\documentclass[12pt,dvips]{article}
\usepackage{epsfig}
\usepackage{feynmf}

\newcommand{\dele}{\mbox{$\Delta E$}}

\newcommand{\raw}{\rightarrow}

\newcommand{\beq}{\begin{equation}}
\newcommand{\eeq}{\end{equation}}
\newcommand{\bqa}{\begin{eqnarray}}
\newcommand{\eqa}{\end{eqnarray}}
\newcommand{\bea}{\begin{array}}
\newcommand{\ena}{\end{array}}

\newcommand{\pip}{\mbox{$\pi^+$}}
\newcommand{\pim}{\mbox{$\pi^-$}}

\newcommand{\piz}{\mbox{$\pi^0$}}

\newcommand{\rhom}{\mbox{$\rho^-$}}
\newcommand{\rhoz}{\mbox{$\rho^0$}}
\newcommand{\et}{\mbox{$\eta$}}
\newcommand{\etpri}{\mbox{$\eta^{'}$}}
\newcommand{\omg}{\mbox{$\omega$}}

\newcommand{\km}{\mbox{$K^-$}}

\newcommand{\dpl}{\mbox{$D^+$}}

\newcommand{\dstz}{\mbox{$D^{*0}$}}

\newcommand{\dstp}{\mbox{$D^{*+}$}}

\newcommand{\dz}{\mbox{$D^0$}}

\newcommand{\bm}{\mbox{$B^-$}}

\newcommand{\bzb}{\mbox{$\bar{B}^0$}}

\begin{document}

\begin{flushright}
{CLNS 97/1503} \\
{CLEO 97-19} \\
{CLEO CONF 97-12} \\
{EPS abstract 338} \\
{To be submitted to PRD} \\

\end{flushright}

\vskip 0.5cm

\begin{center}
{\bf  Search for Color-Suppressed $B$ Hadronic Decay}

{\bf  Processes with CLEO}
\end{center}

\vskip 1.0cm

\begin{center}
{\large  CLEO Collaboration}

\vskip 0.6cm

{\large  (August 7, 1997)}
\end{center}

\vskip 2.0cm

\begin{center}
{\Large  Abstract}
\end{center}

        Using 3.1 fb$^{-1}$ of data accumulated at the $\Upsilon$(4S) by the
CLEO-II detector, corresponding to 3.3 million $\mbox{$B\bar{B}$}$
pairs, we have searched for the color-suppressed $B$ 
hadronic decay processes: $\bzb\raw\dz(\dstz)$X$^0$, where X$^0$ is 
a light neutral meson $\piz$, $\rhoz$, $\et$, $\etpri$ or $\omg$.
The $\dstz$ mesons are reconstructed in $\dstz \raw \dz\piz$ and the $\dz$
mesons in 
$\dz \raw \km\pip$, $\km\pip\piz$ and $\km\pip\pip\pim$ decay modes.
No obvious signal is observed.  We set 90\% C.L. upper limits  
on these modes, varying from 1.2 $\times10^{-4}$ for $\bzb\raw\dz\piz$ to 
1.9 $\times10^{-3}$ for $\bzb\raw\dstz\etpri$.


\vskip 0.5cm


\newpage
\begin{center}
B.~Nemati,$^{1}$ S.~J.~Richichi,$^{1}$ W.~R.~Ross,$^{1}$
P.~Skubic,$^{1}$
M.~Bishai,$^{2}$ J.~Fast,$^{2}$ J.~W.~Hinson,$^{2}$
N.~Menon,$^{2}$ D.~H.~Miller,$^{2}$ E.~I.~Shibata,$^{2}$
% ...
\end{center}

\small
\begin{center}
$^{1}${University of Oklahoma, Norman, Oklahoma 73019}\\
$^{2}${Purdue University, West Lafayette, Indiana 47907}\\
% ...
\end{center}

\newpage

\vskip 0.5cm

\begin{center}
{\bf  I. INTRODUCTION}
\end{center}

\begin{figure}[hhh]
   \begin{center}
   \leavevmode
   \epsfxsize=8.5cm
   \epsfbox{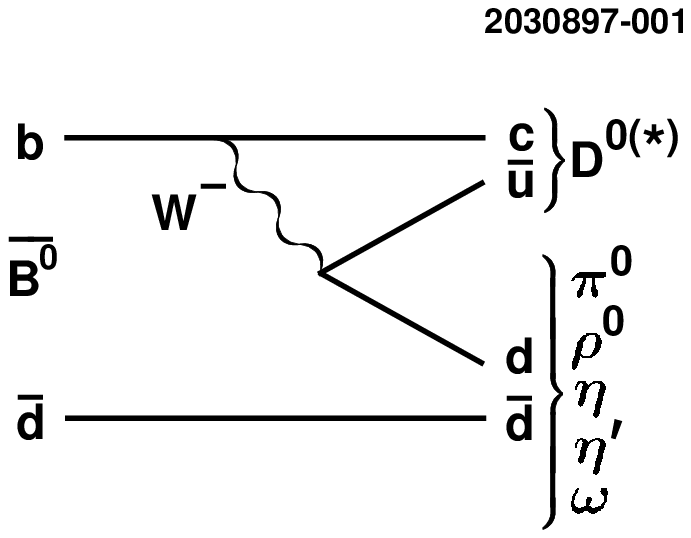}
   \end{center}
   \caption {Internal spectator diagram of $B$ hadronic  
             decays $\bzb\raw\dz(\dstz)$X$^0$.}
\label{fg01}
\end{figure}

The $B$ hadronic decays $\bzb\raw\dz(\dstz)$X$^0$, where X$^0$ is 
a light neutral meson  $\piz$, $\rhoz$, $\et$, $\etpri$ or $\omg$, 
have not yet been observed. These decays proceed via the internal 
spectator diagram shown in Figure ~\ref{fg01}.
The internal spectator decays are expected to be suppressed relative 
to the decays that proceed via external spectator diagrams, since  
the color of the quarks from the virtual W must match 
the color of the c quark and the accompanying spectator antiquark.
Therefore these decays are refered to as color-suppressed 
decays, while decays via external spectator diagrams are refered
to as color-favored decays. 
Measurements of these color-suppressed decays allow tests of the 
factorization~\cite{factor} hypothesis  
and provide useful information on the scale of strong final-state interaction 
in the $B$ meson system.

Previous CLEO papers~\cite{bigb} reported upper limits on these 
color-suppressed $B$ hadronic decays.  Here we present 
new results 
using the full CLEO-II data set and an improved analysis method.

\vskip 0.5cm

\begin{center}
{\bf  II. DATA SAMPLE and EVENT SELECTION}
\end{center}

The data used in this analysis were produced in $e^+e^-$ annihilations at the
Cornell Electron Storage Ring (CESR) and collected with the 
CLEO-II detector \cite{detector}. 
The intergrated luminosity is 3.1 fb$^{-1}$ at the $\Upsilon$(4S) resonance,
which corresponds to (3.32$\pm$0.07) $\times$ 10$^6$ 
$\mbox{$B\bar{B}$}$ pairs, 
and 1.6 fb$^{-1}$ at energies just below $\mbox{$B\bar{B}$}$ threshold  
(henceforth referred to as the continuum).

Hadronic events are selected by requiring at least three charged tracks, 
a total detected energy of at least 0.15 E$_{c.m}$, and a primary vertex 
within 5.0 cm along the beam ($z$) axis of the interaction point.
To suppress continuum background, we require that the ratio of second
to zeroth Fox-Wolfram moments R$_2$ \cite{r2} determined using charged 
tracks and unmatched neutral showers be less than 0.3 (0.5 for clean decay 
modes involving $\et$ or $\etpri$).
To further reduce continuum
background, we then require that the cosine of the angle between the
sphericity axis of the $B$ meson candidate and the sphericity axis of the 
remainder of the event satisfies $|$cos($\theta_{sphericity}$)$|$ $<$ 0.8
(0.9 for decay modes involving $\et$ or $\etpri$). 
For a jet-like continuum event, the two axes are almost parallel,
while they are almost uncorrelated for a $B\bar{B}$ event, 

\vskip 0.5cm
\pagebreak

\begin{center}
{\bf  III. $B$ RECONSTRUCTION}
\end{center}

\vskip 0.5cm

\begin{center}
{\bf Selection of  $\dz$ and $\dstz$ candidates}
\end{center}

The $\dz$ candidates are
reconstructed in the decay modes $\dz\raw\km\pip$, $\km\pip\piz$ and 
$\km\pip\pip\pim$ (charge-conjugate modes are implied). 
The $\piz$ candidates are formed by combining two showers 
whose invariant mass is within 2.5$\sigma$ of the $\piz$ mass (where
henceforth $\sigma$ denotes  r.m.s. resolution). 
Charged tracks are required to be consistent with coming from the
interaction region in both the $r -  \phi$ and $r - z$ planes.
The measured specific inoization (dE/dx) of charged kaon and pion
candidates are required to be consistent to within 2$\sigma$ 
for kaon candidates and 3$\sigma$ for pion candidates.
Charged tracks are required to have a momentum greater than
250 MeV for $\dz\raw\km\pip$ and $\dz\raw\km\pip\piz$ candidates
and 200 MeV for $\dz\raw\km\pip\pip\pim$ candidates.
For $\dz\raw\km\pip\piz$ decay mode, we select regions of the Dalitz
plot with large amplitude to further suppress the combinatoric backgrounds. 
The invariant mass of $\dz$ candidates is required to be within 2.0$\sigma$
of the nominal $\dz$ mass.

The $\dstz$ candidates are reconstructed using decay mode $\dstz\raw\dz\piz$. 
We form $\dstz$ candidates by $\dz$ candidates using above selection, then 
require that the $\dstz$ $-$ $\dz$ mass difference be within 2.5$\sigma$ 
of the measured value \cite{PDG}. 

\vskip 0.5cm

\begin{center}
{\bf Selection of the light neutral meson X$^0$}
\end{center}

We reconstruct $\piz$ candidates as described previously.  The $\rhoz$
candidates are reconstructed in the mode $\rhoz\raw\pip\pim$.  

Candidate $\et$ and $\etpri$ mesons are reconstructed in their 
$\et\raw\gamma\gamma$ and $\etpri\raw\et\pip\pim$ decay modes. The absolute
value of the $\et$ decay angle is required to be less than 0.85 to remove
asymmetric candidates which are primarily background.  The invariant mass 
of each $\et$ and $\etpri$
candidate must be within 30 MeV of their nominal mass.

The $\omg$ mesons are reconstructed in the decay mode $\omg\raw\pip\pim\piz$.
Charged and neutral pions are required to have momenta greater than 250 MeV, 
to reject soft pions from $\dstz$ or $\dstp$ decays. The $\omg$ candidates are 
also required to be within 30 MeV of the nominal $\omg$ mass.

All charged pion candidates used in X$^0$ reconstruction are
required to have a measured dE/dx within 3$\sigma$ of the
expected value for pions.

\vskip 0.5cm

\begin{center}
{\bf Selection of the $B$ candidates}
\end{center}

The $D^{(*)0}$ candidates are combined with a light X$^0$ to form a B meson.
At CLEO the energy of the $B$ meson is the same as the beam energy and the
measured beam energy is more precise than the reconstructed $B$ meson energy.
Full reconstruction of $B$ mesons at CLEO makes use of this fact
by defining two variables.
One is the beam-constrained mass, 
M$_B \equiv \sqrt{E^2_{beam} - P^{2}_{observed}}$.
The other one is the difference between the reconstructed energy and the 
beam energy, 
$\dele \equiv E_{observed}$ $-$ $E_{beam}$.
The $\dele$ variable is sensitive to missing or extra particles in the 
$B$ decay, 
as well as particle species. 
For fully-reconstructed $B$ meson decays, the M$_B$ distribution peaks at 
5.28 GeV with resolution around 2.7 MeV, and $\dele$ peaks at 0.0 GeV with 
a resolution ranging from 18 to 50 MeV, depending on the $B$ and $\dz$ decay
modes.

Since signal and background are in general much better seperated in 
$\dele$ than in M$_B$, instead of cutting on the $\dele$ variable and 
fitting M$_B$ as in previous analyses, we cut on M$_B$ and fit the $\dele$ 
distribution for the signal yield.

\vskip 0.5cm
\pagebreak

\begin{center}
{\bf  IV. BACKGROUND STUDY}
\end{center}

In our search for the color-suppressed $B$ hadronic decay modes
$\bzb\raw D^{(*)0}(\piz, \rhoz, \et, \etpri, \omg)$,
there are backgrounds to these decays from continuum and 
$\mbox{$B\bar{B}$}$ events.
The continuum backgrounds are suppressed using event-shape variables.
They are not expected to show any structure in the $\dele$ distributions.  
The 1.6 fb$^{-1}$ continuum data set is used to monitor the continuum
background levels.  
We find the continuum background level to be very low for
all color-suppressed modes. 
No accumulation around $\dele$ $=$ 0 is observed in the continuum data.

The backgrounds from $\mbox{$B\bar{B}$}$ events are
dominated by feedthrough from color-favored two-body hadronic decays of 
the type:
\begin{center}
$\bm\raw\dz(\pim,\rhom,a_{1}^{-})$, $\bm\raw\dstz(\pim,\rhom,a_{1}^{-})$

$\bzb\raw\dpl(\pim,\rhom,a_{1}^{-})$, $\bzb\raw\dstp(\pim,\rhom,a_{1}^{-})$
\end{center}
The branching ratios of these color-favored $B$ meson decay modes were
measured previously by CLEO \cite{bigb}.
In most cases the background arises when a real, energetic $\dz$ or 
$\dstz$ from the 2-body color-favored decays is combined with a fake 
light meson. 

The backgrounds from these color-favored processes can have
structure in the M$_B$ and $\dele$ distributions, depending on
which color-suppressed mode is being analyzed.  Particularly important
are color-favored B meson decays that give exactly the same final state 
particles as our 
color-suppressed signals do. Neither misidentification nor additional
particles are needed for those color-favored decays to fake some signal modes. 
Therefore,
the M$_B$ distribution from these physics background peaks at 5.28 GeV 
while its $\dele$ distribution peaks at 0.0 GeV, exactly as the 
color-suppressed signal.  While $\dz\piz$ is not susceptible to this
background $\dz\rhoz$ and $\dz\omega$ are, as shown below.

\[
\begin{array}{lrr}
{\rm Color-suppressed:} & \bzb\raw\dz\rhoz\raw\dz\pip\pim & 
\bzb\raw\dz\omg\raw\dz\pip\pim\piz\\
{\rm Color-favored:}    & \bzb\raw\dstp\pim\raw\dz\pip\pim & 
\bzb\raw\dstp\rhom\raw\dz\pip\pim\piz\\
\end{array} 
\]

Another background that can show structure is color-favored decay
in which one of the final state particles is lost.  Examples include:
\[
\begin{array}{lr@{\hspace{.25em}}lr@{\hspace{.25em}}l}
{\rm Color-suppressed:} & \bzb\raw\dz\piz & & \bzb\raw\dz\rhoz\raw\dz\pip\pim 
& \\
{\rm Color-favored:}& \bm\raw\dz\rhom\raw\dz\piz & (\pim) & 
\bzb\raw\dstp\rhom\raw\dz\pip\pim & (\piz)  \\
\end{array} 
\]
These background events can peak in M$_B$ around 5.28 GeV
when the missing $\pim$ or $\piz$ from the $\rhom$ decay is very soft
and does not contribute much to the beam-constrained mass calculation. 
However, the $\dele$ for these background events differs from zero by 
more than one pion mass, due to the missing $\pim$ or $\piz$ 
from the $\rhom$ decay.
For these types of color-favored backgrounds, the 
color-suppressed signals are much better separated from background in $\dele$.

For decay modes involving $\et$ or $\etpri$, 
combinatoric background is the dominating source. 
Therefore, backgrounds for these color-suppressed processes 
have no accumulation in the M$_B$ and $\dele$ distributions.

For $\bzb\raw\dstz$X$^0$, there is no corresponding color-favored $B$ meson
decay that fakes our signal as $\bzb\raw \dstp\pim$ fakes
$\bzb\raw\dz\rhoz$. Also the background level from color-favored $B$ meson 
decays is very low for $\bzb\raw\dstz$X$^0$ decay processes,
due to the good resolution on the $\dstz-\dz$ mass difference.

Almost all the discrimination power against color-favored physics backgrounds
come from selection cuts on X$^0$. We make full use of mass, momentum,
decay angle and other kinematic variables of X$^0$ to suppress backgrounds 
while keeping signal efficiency as high as possible.

The X$^0$ candidates in $\bzb\raw D^{(*)0}{\rm X}^0$ are very energetic
due to the hard spectrum of two-body $B$ meson decays.
We require the momentum
of the $\piz$ candidate to range from 2.1 GeV to 2.5 GeV.
Similar momentum requirements are imposed on the other light 
neutral meson X$^0$ candidates.

For $\bzb\raw\dz\rhoz$ decays, there are color-favored physics
backgrounds from $\bzb\raw\dstp\pim$ that give exactly the same final
state particles. The $\bm\raw\dz\rhom$ decay can also fake 
our color-suppressed signal by substituting the soft $\piz$ from $\rhom$ 
decay by a soft $\pip$ from the other $B$ meson. 
In these physics backgrounds, the $\pim$ is always much more energetic
than the $\pip$ from $\dstp\raw\dz\pip$ decay. There exists a correlation 
between the $\dz$ and the fast $\pim$ (slow $\pip$) from the fake $\rhoz$. 
To suppress these physics backgrounds, we require that the $\dz$ candidate 
to be  associated with a fast $\pip$ (slow $\pim$) from the $\rhoz$ candidate. 
There is still a contribution from color-favored physics backgrounds even
after this requirement, because a $\dz$ decay has a certain chance of being
misidentified as a $\bar{\dz}$ decay. 
For the $\dz$'s from our signal process, together with the dE/dx and 
$\dz$ mass requirements, this misidentification rate is determined to be 
less than 20\%. 
After further suppression due to the $\rhoz$ mass and momentum requirements,
the contribution from color-favored physics background is negligible.
Since the $\rhoz$ from $\bzb\raw\dz\rhoz$ decay is longitudinally polarized, 
we also cut on the $\rhoz$ decay angle (the angle between the direction of 
the pion in the $\rhoz$ rest frame and the direction of the $\rhoz$ in 
the lab frame) to reduce combinatoric backgrounds.

Signal selection efficiencies for all the color-suppressed decay modes are
shown in Table 1. The systematic error due to the detection of charged 
and neutral tracks, together with the Monte Carlo statistical error, 
are included in the error on the efficiency for each decay mode.

\begin{table}[hhh]
\caption{Selection efficiencies and yields of all color-suppressed modes.
The three efficiencies and yields of each 
$\bzb\raw\dz(\dstz)$X$^0$ correspond to the three 
$\dz\raw\km\pip$, $\km\pip\piz$ and $\km\pip\pip\pim$ modes.}

 \vskip 0.5cm
 
 \begin{center}
 \begin{tabular}{|l|c|c|c|} \hline
     Decay Mode  &  Selection Efficiency  & Yield  \\ \hline
 \hline
$\bzb\raw\dz\piz$     & 26.1$\pm$2.2,  7.8$\pm$1.0,  12.5$\pm$1.3\% 
                      & -0.3$\pm$6.4, -6.7$\pm$4.3,  -3.3$\pm$7.0   \\ \hline
$\bzb\raw\dstz\piz$   & 14.1$\pm$1.8,  3.7$\pm$0.7,   5.4$\pm$0.9\% 
                      &  2.5$\pm$2.6,  5.0 $\pm$3.4, -1.2$\pm$3.4   \\ \hline
$\bzb\raw\dz\rhoz$    &  8.4$\pm$0.4,  2.6$\pm$0.3,   3.9$\pm$0.3\% 
                      &  1.4$\pm$3.0, -3.0$\pm$4.3,   3.1$\pm$5.0   \\ \hline
$\bzb\raw\dstz\rhoz$  &  4.0$\pm$0.4,  1.0$\pm$0.2,   1.5$\pm$0.2\% 
                      & -1.0$\pm$1.4,  1.4$\pm$1.6,   0.8$\pm$1.3   \\ \hline
$\bzb\raw\dz\et$      & 24.5$\pm$3.0,  7.0$\pm$1.2,  11.4$\pm$1.6\% 
                      & -1.4$\pm$2.0, -3.1$\pm$3.1,  -6.0$\pm$4.0   \\ \hline
$\bzb\raw\dstz\et$    & 10.5$\pm$1.8,  3.4$\pm$0.8,   4.9$\pm$0.9\% 
                      &       0,           0,              0        \\ \hline
$\bzb\raw\dz\etpri$   & 13.4$\pm$1.9,  3.6$\pm$0.7,   5.9$\pm$1.0\% 
                      &       0,       0.8$\pm$2.2,   1.8$\pm$3.0   \\ \hline
$\bzb\raw\dstz\etpri$ &  5.9$\pm$1.1,  1.7$\pm$0.4,   2.5$\pm$0.5\% 
                      &       0,           0,              1        \\ \hline
$\bzb\raw\dz\omg$     & 12.4$\pm$1.3,  2.8,$\pm$0.4,  3.2$\pm$0.5\% 
                      & -4.1$\pm$4.0,  6.2$\pm$3.8,   3.6$\pm$5.6   \\ \hline
$\bzb\raw\dstz\omg$   &  4.8$\pm$0.7,  1.0$\pm$0.2,   1.3$\pm$0.2\% 
                      &  1.8$\pm$1.2,  0.8 $\pm$1.8, -0.2$\pm$1.2   \\ \hline
 \hline
\end{tabular}
\end{center}
\end{table}

\vskip 0.5cm

\begin{center}
{\bf  V. RESULTS}
\end{center}

The $\dele$ distributions 
for the on $\Upsilon$(4S) and continuum data samples
of all the color-suppressed signal processes
after all cuts are shown in Fig. 2 $-$ 6.
The $\dele$ distribution of each color-suppressed mode is fit
with a Gaussian and a background shape function.
The mean value and width of the 
Gaussian distribution are fixed with values determined from
signal Monte Carlo. We use various color-favored decay modes
$\bm\raw\dz\pim$, $\bm\raw\dz\rhom$, $\bm\raw\dstz\pim$, $\bm\raw\dstz\rhom$,
$\bzb\raw\dpl\pim$, $\bzb\raw\dpl\rhom$ to check that the  $\dele$ 
resolutions in data and Monte Carlo agree well. Possible differences between 
data and Monte Carlo in the $\dele$ distributions are considered and included 
in the yield error as systematic errors.

Various $\dele$ background shape functions have
been used to fit for the signal yield: a simple second-order polynomial;
or a background shape using Monte Carlo simulation 
$\mbox{$B\bar{B}$}$ 
events plus a continuum component represented by a second-order polynomial.
For the latter shape, the $\mbox{$B\bar{B}$}$ contribution is scaled 
to the known luminosity while the continuum component is allowed to float.
Our results are found to
be insensitive to different background shapes, and both of them describe
the $\dele$ distributions reasonably well. Differences in the yield due to
the choice of $\dele$ background shape are included in the yield error
to account for the systematic uncertainties.
For each signal process with several
$\dz$ decay submodes, the yield for each $\dz$ submode is obtained seperately,
since the $\dele$ resolutions are different for the different modes.
The results are shown in Table 1.
The yields of the $\dz$ submodes are added independently to get the 
total yield.

The formulas used to calculate the branching fractions are:
\begin{equation}
Br(\bzb\raw\dz X^{0})=\frac{N_{obs}}{N_{B\bar{B}}\times
(\sum_{i=1}^3 {\rm Efficiency}(i)\times Br(\dz_{i}))\times \prod Br(X^{0})}
\end{equation}

\begin{equation}
Br(\bzb\raw\dstz X^{0})=\frac{N_{obs}}{N_{B\bar{B}}\times 
Br(\dstz\raw\dz\piz) \times
(\sum_{i=1}^3 {\rm Efficiency}(i)\times Br(\dz_{i}))\times \prod Br(X^{0})}
\end{equation}
where N$_{obs}$ is the total yield summed over the three $\dz$ submodes,
N$_{B\bar{B}}$ is the number of $\mbox{$B\bar{B}$}$ pairs,
Efficiency(i) is the selection efficiency for 
$\bzb\raw\dz(\dstz)X^{0}$ decay in the {\it i}th $\dz$ submode, 
$Br(\dz_{i})$ is the branching ratio of the {\it i}th $\dz$ decay mode, 
and $\prod Br(X^{0})$ is the product
over all the relevant branching fractions of the X$^0$ decay chain.
Particle Data Group values for $\dz$, $\dstz$, $\et$, $\etpri$ and $\omg$ 
branching ratios are used in the upper limits calculation \cite{PDG} and
are listed in Table 2.

\begin{table}[hhh]
\caption{Particle Data Group branching ratios that are used in the
upper limit calculation for color-suppressed B hadronic decays.}

 \vskip 0.5cm
 
 \begin{center}
 \begin{tabular}{|l|c|c|c|} \hline
     Decay Mode  &  PDG Branching Ratio \\ \hline
 \hline
$\dz\raw\km\pip$         &    4.01$\pm$0.14\%   \\ \hline
$\dz\raw\km\pip\piz$     &    13.8$\pm$1.0\%    \\ \hline
$\dz\raw\km\pip\pip\pim$ &     8.1$\pm$0.5\%    \\ \hline
$\dstz\raw\dz\piz$       &    63.6$\pm$2.8\%    \\ \hline
$\rhoz\raw\pip\pim$      &      100\%           \\ \hline
$\et\raw\gamma\gamma$    &    38.8$\pm$0.5\%    \\ \hline
$\etpri\raw\et\pip\pim$  &    43.7$\pm$1.5\%    \\ \hline
$\omg\raw\pip\pim\piz$   &    88.8$\pm$0.7\%    \\ \hline
 \hline
\end{tabular}
\end{center}
\end{table}

The upper limits of color-suppressed branching ratios are determined by the 
method described in section 17 of the Particle Data Group \cite{PDG}. 
90\% C.L. upper limits on branching ratios of color-suppressed $B$ hadronic 
decay processes, together with theoretical predictions
\cite {model1}, are shown in Table 3.
Among all the decay modes, the upper limit for the $\bzb\raw\dz\piz$ mode is 
the lowest at 1.2 $\times10^{-4}$. All the upper limits on branching
ratios are still higher than theoretical predictions \cite {model1,model2}. 
Compared with factorization and QCD based calculations,
no dramatic enhancement of color-suppressed $B$ hadronic decay branching
ratios is observed, indicating that there is no sign of large scale 
final-state interaction in these $B$ meson decay modes.

\begin{table}[hhh]
\caption{90\% C.L. upper limits in branching ratios of all color-suppressed 
modes, together with comparison with theoretical predictions.}

 \vskip 0.5cm
 
 \begin{center}
 \begin{tabular}{|l|c|c|} \hline
     Decay Mode  &  Branching Ratio (@90\% C.L.)  & Theoretical Predictions 
\\ \hline
 \hline
$\bzb\raw\dz\piz$     & $<$ 1.2$\times10^{-4}$  & 0.7$\times10^{-4}$ \\ \hline
$\bzb\raw\dstz\piz$   & $<$ 4.4$\times10^{-4}$  & 1.0$\times10^{-4}$ \\ \hline
$\bzb\raw\dz\rhoz$    & $<$ 3.9$\times10^{-4}$  & 0.7$\times10^{-4}$ \\ \hline
$\bzb\raw\dstz\rhoz$  & $<$ 5.6$\times10^{-4}$  & 1.7$\times10^{-4}$ \\ \hline
$\bzb\raw\dz\et$      & $<$ 1.3$\times10^{-4}$  & 0.5$\times10^{-4}$ \\ \hline
$\bzb\raw\dstz\et$    & $<$ 2.6$\times10^{-4}$  & 0.6$\times10^{-4}$ \\ \hline
$\bzb\raw\dz\etpri$   & $<$ 9.4$\times10^{-4}$  &                    \\ \hline
$\bzb\raw\dstz\etpri$ & $<$  19$\times10^{-4}$  &                    \\ \hline
$\bzb\raw\dz\omg$     & $<$ 5.1$\times10^{-4}$  & 0.7$\times10^{-4}$ \\ \hline
$\bzb\raw\dstz\omg$   & $<$ 7.4$\times10^{-4}$  & 1.7$\times10^{-4}$ \\ \hline
 \hline
\end{tabular}
\end{center}
\end{table}

\centerline{\bf ACKNOWLEDGEMENTS}
\smallskip
We gratefully acknowledge the effort of the CESR staff in providing us with
excellent luminosity and running conditions.
J.P.A., J.R.P., and I.P.J.S. thank
the NYI program of the NSF,
M.S. thanks the PFF program of the NSF,
G.E. thanks the Heisenberg Foundation,
K.K.G., M.S., H.N.N., T.S., and H.Y. thank the
OJI program of DOE,
J.R.P., K.H., M.S. and V.S. thank the A.P. Sloan Foundation,
R.W. thanks the
Alexander von Humboldt Stiftung, 
and M.S. thanks Research Corporation
for support.
This work was supported by the National Science Foundation, the
U.S. Department of Energy, and the Natural Sciences and Engineering Research
Council of Canada.

\begin{figure}[hhh]
   \begin{center}
   \leavevmode
   \epsfxsize=17.5cm
   \epsfbox{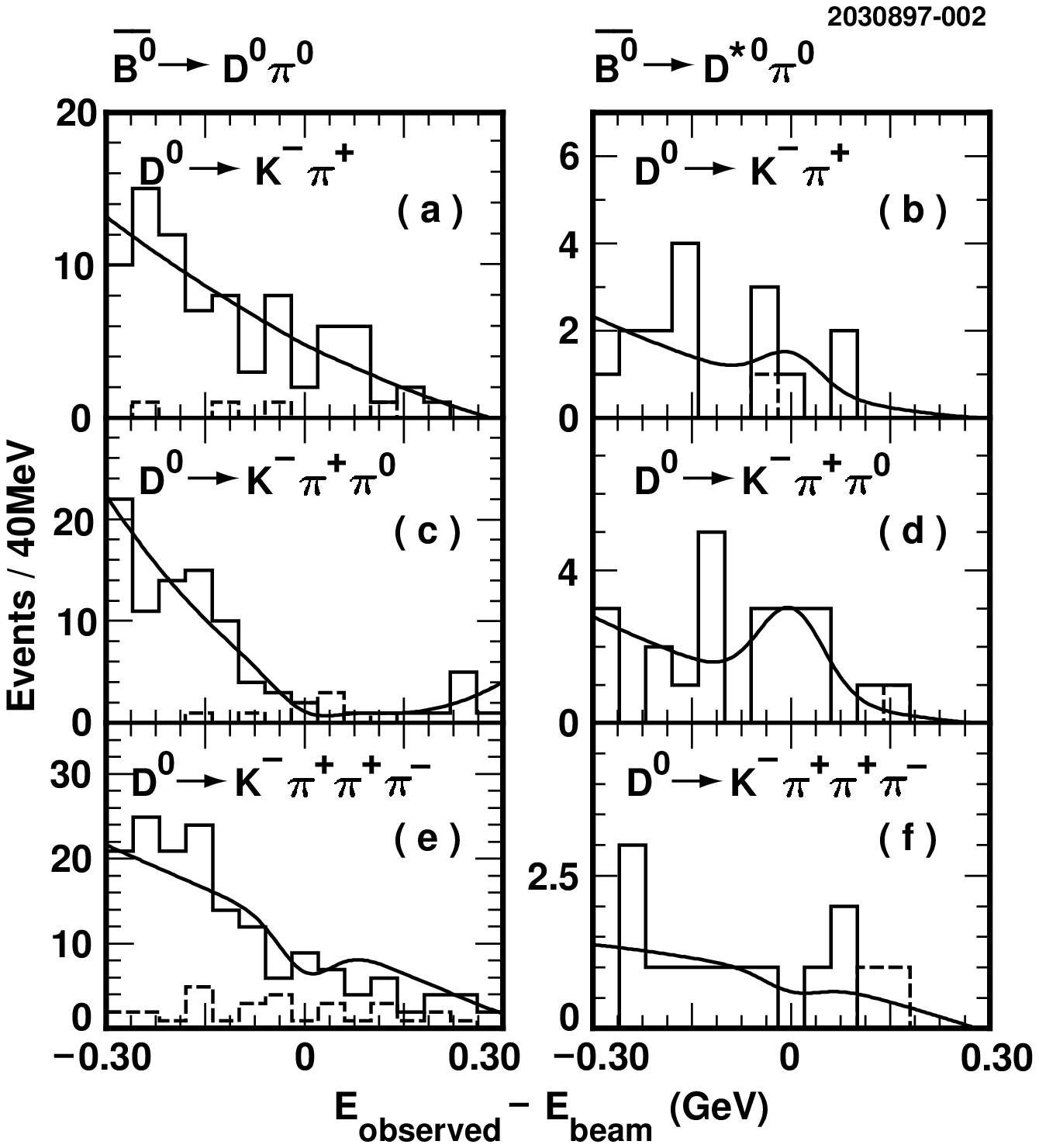}
   \end{center}
   \caption {$\dele$ distributions of $\bzb\raw\dz\piz$ and
  $\bzb\raw\dstz\piz$ decay modes. Solid histograms are the $\dele$ 
  distributions of the 3.1 fb$^{-1}$ of data collected on the $\Upsilon$(4S) 
  resonance, which are fit using background and signal functions. 
  Dashed histograms are from the 1.6 fb$^{-1}$ continuum data sample. }
\label{fg02}
\end{figure} 

\begin{figure}[hhh]
   \begin{center}
   \leavevmode
   \epsfxsize=17.5cm
   \epsfbox{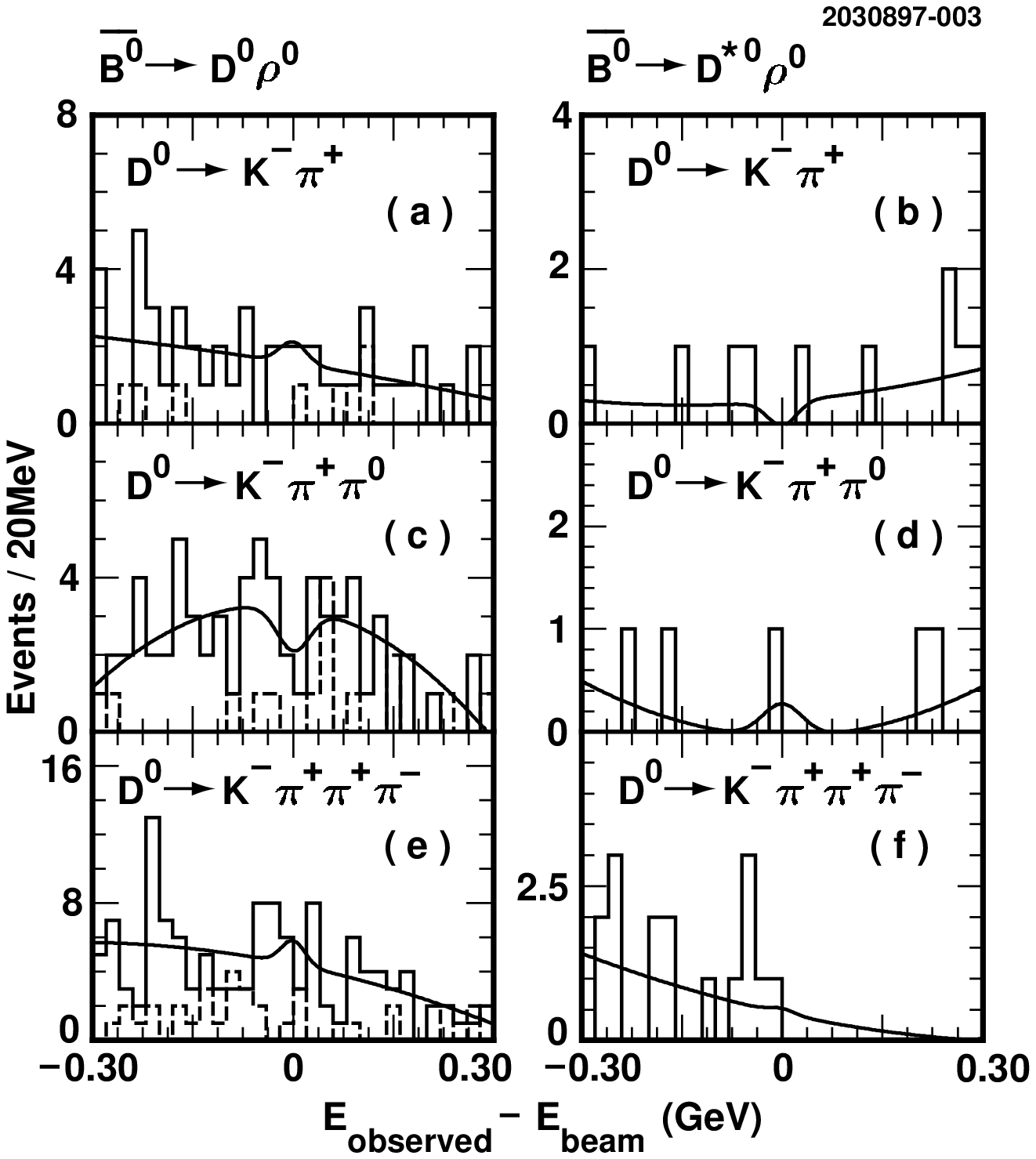}
   \end{center}
   \caption {$\dele$ distributions of $\bzb\raw\dz\rhoz$ and
  $\bzb\raw\dstz\rhoz$ decay modes. Solid histograms are the $\dele$ 
  distributions of the 3.1 fb$^{-1}$ of data collected on the $\Upsilon$(4S) 
  resonance, which are fit using background and signal functions. 
  Dashed histograms are from the 1.6 fb$^{-1}$ continuum data sample. }
\label{fg03}
\end{figure} 

\begin{figure}[hhh]
   \begin{center}
   \leavevmode
   \epsfxsize=17.5cm
   \epsfbox{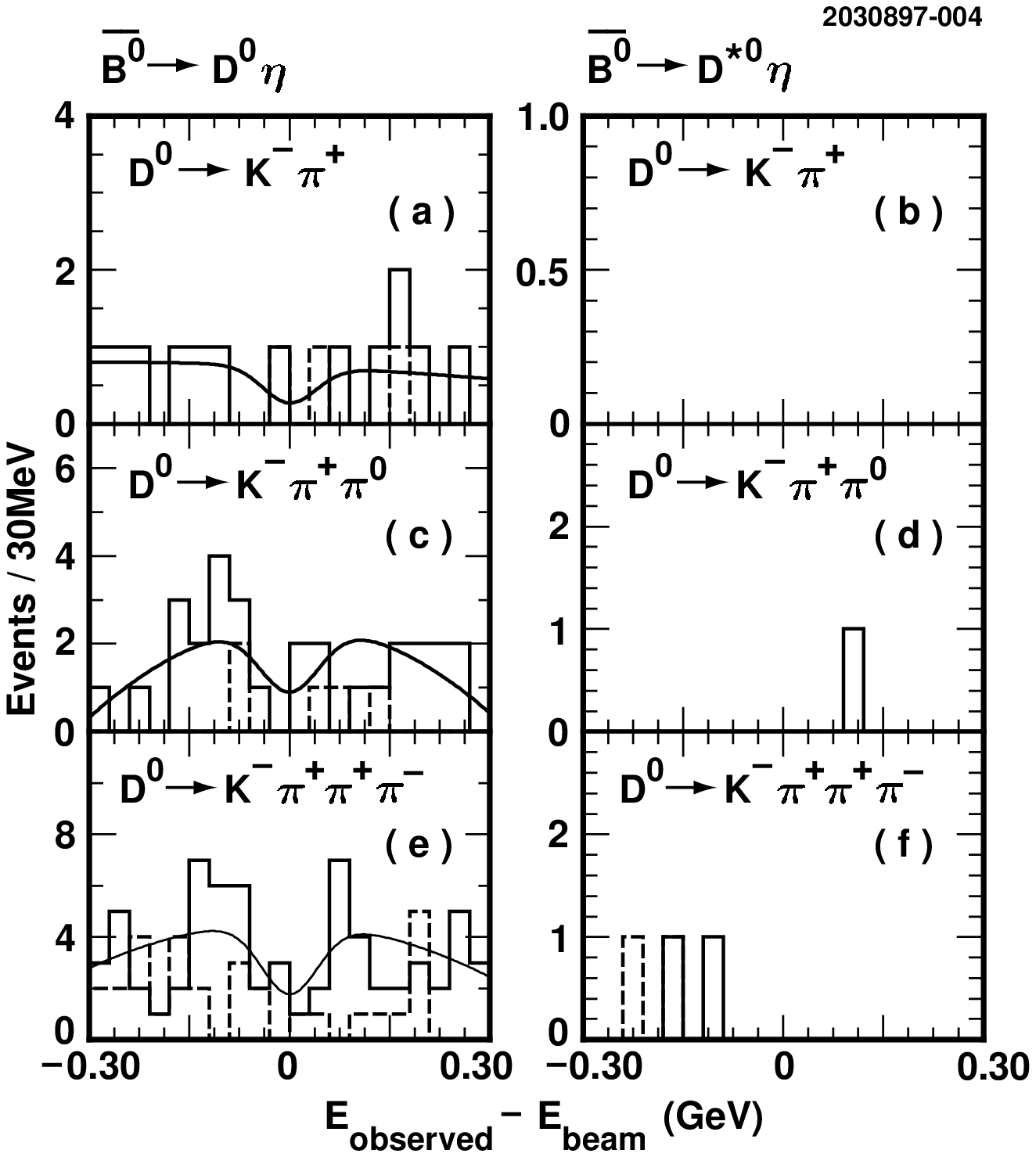}
   \end{center}
   \caption {$\dele$ distributions of $\bzb\raw\dz\et$ and
  $\bzb\raw\dstz\et$ decay modes. Solid histograms are the $\dele$ 
  distributions of the 3.1 fb$^{-1}$ of data collected on the $\Upsilon$(4S) 
  resonance, which are fit using background and signal functions. 
  Dashed histograms are from the 1.6 fb$^{-1}$ continuum data sample. }
\label{fg04}
\end{figure}

\begin{figure}[hhh]
   \begin{center}
   \leavevmode
   \epsfxsize=17.5cm
   \epsfbox{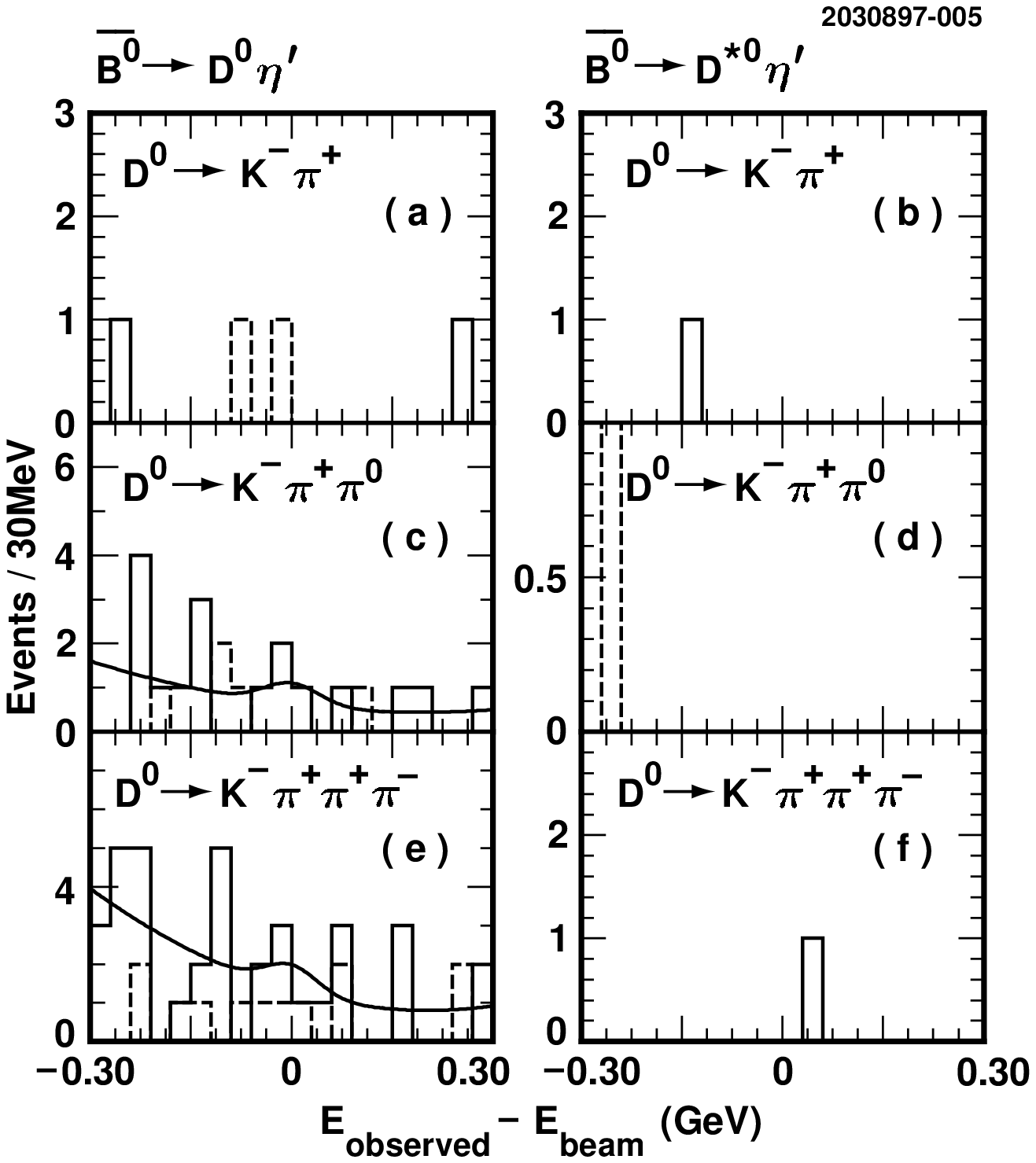}
   \end{center}
   \caption {$\dele$ distributions of $\bzb\raw\dz\etpri$ and
  $\bzb\raw\dstz\etpri$ decay modes. Solid histograms are the $\dele$ 
  distributions of the 3.1 fb$^{-1}$ of data collected on the $\Upsilon$(4S) 
  resonance, which are fit using background and signal functions. 
  Dashed histograms are from the 1.6 fb$^{-1}$ continuum data sample. }
\label{fg05}
\end{figure}

\begin{figure}[hhh]
   \begin{center}
   \leavevmode
   \epsfxsize=17.5cm
   \epsfbox{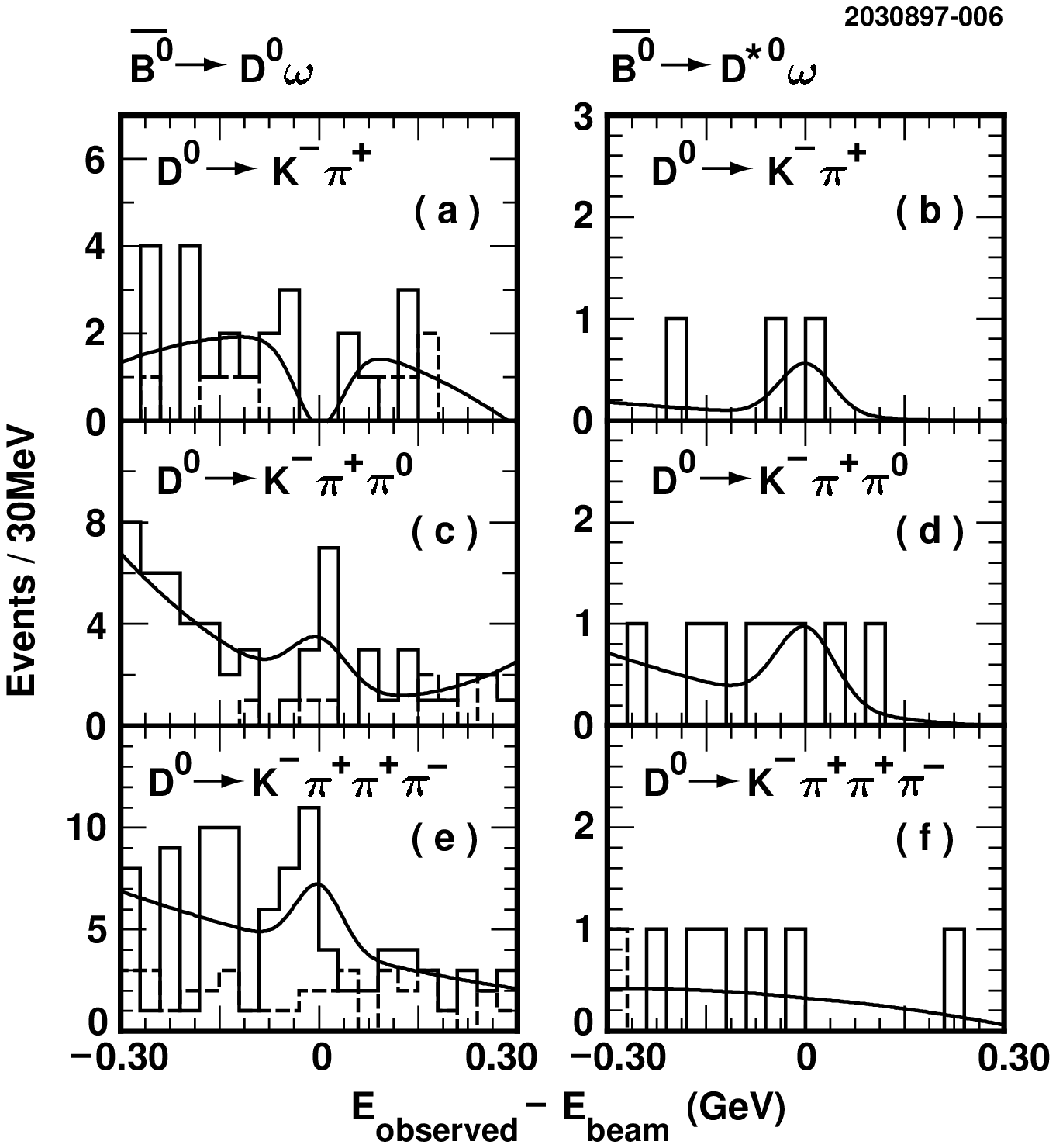}
   \end{center}
   \caption {$\dele$ distributions of $\bzb\raw\dz\omg$ and
  $\bzb\raw\dstz\omg$ decay modes. Solid histograms are the $\dele$ 
  distributions of the 3.1 fb$^{-1}$ of data collected on the $\Upsilon$(4S) 
  resonance, which are fit using background and signal functions. 
  Dashed histograms are from the 1.6 fb$^{-1}$ continuum data sample. }
\label{fg06}
\end{figure}


\end{document}